\documentclass[universe,article,accept,moreauthors,pdftex,10pt,a4paper]{Definitions/mdpi}
\firstpage{1} 
\pubvolume{10}
\issuenum{159}
\articlenumber{159}
\pubyear{2024}
\copyrightyear{2024}
\externaleditor{Academic Editor: Chandrachur Chakraborty and Banibrata Mukhopadhyay}
\datereceived{31 October 2022} 
\daterevised{27 February 2024} 
\dateaccepted{19 March 2024} 
\datepublished{28 March 2024} 
\hreflink{https://doi.org/10.3390/ universe10040159} 
 
\pdfoutput=1

\Title{\texorpdfstring{Boyer--Lindquist  Space-Times and Beyond: Metamaterial Analogues for Arbitrary
Space-Times}
{Boyer--Lindquist  space-times and beyond: Metamaterial analogues for arbitrary space-times}}

\TitleCitation{Boyer--Lindquist  Space-Times and Beyond: Metamaterial Analogues for Arbitrary Space-Times}

\Author{Sebastian Schuster  $^{1,2}$$^{,\dagger}$\orcidA{} and Matt Visser ${^{3,}}$*$^{,\dagger}$\orcidB{}}

\AuthorNames{Sebastian Schuster and Matt Visser}

\AuthorCitation{Schuster, S.; Visser, M.}

\address{%
$^{1}$ \quad \'Ustav Teoretick\'e Fyziky, Matematicko-Fyzik\'aln\'i Fakulta, Univerzita Karlova, V Hole\v{s}ovi\v{c}k\'ach~2, \linebreak  180~00
	{ Praha,} 
 Czech Republic; sebastian.schuster@utf.mff.cuni.cz\\
$^{2}$ \quad
School of Mathematics and Statistics, University of Sheffield, Hounsfield Road, Sheffield S3~7RH,  UK\\
$^{3}$ \quad
School of Mathematics and Statistics,
Victoria University of Wellington, P.O.~Box~600, Wellington~6140, 
New~Zealand
}

\corres{Correspondence: matt.visser@sms.vuw.ac.nz}

\firstnote{These authors contributed equally to this work.} 

\abstract{%
Analogue space-times (and in particular metamaterial analogue space-times) have a long {varied} and {rather} complex history. Much of the {previous} related work {to this field} has focused on spherically symmetric models; however, axial symmetry is much more relevant for {mimicking astrophysically interesting} systems {that are typically} subject to rotation.
{Now it is well known that} physically reasonable stationary axisymmetric space-times can, under very mild technical conditions, 
be put into Boyer--Lindquist  form. 
Unfortunately, a metric presented in Boyer--Lindquist  form is not well adapted to the ``quasi-Cartesian'' 
metamaterial analysis {that} we developed in our previous articles on ``bespoke  analogue space-times''.
In the current article, we shall first focus specifically on {various} space-time metrics presented in Boyer--Lindquist  form, and {subsequently} determine {a suitable set of} equivalent metamaterial susceptibility tensors in a laboratory setting. 
We {shall} then turn to analyzing generic space-times, not even necessarily stationary, again determining {a suitable set of} equivalent metamaterial susceptibility tensors. 
Perhaps surprisingly, we find that the {well-known} ADM formalism {proves to be} not particularly useful, and that {it is instead} the {dual} ``threaded'' (Kaluza--Klein--inspired) formalism {that} provides {much} more tractable results.
While the background laboratory metric is {(for mathematical simplicity and physical plausibility)} always taken to be Riemann flat, we {will} allow for arbitrary curvilinear coordinate systems {on the flat background space-time}.  Finally, {for completeness}, we {shall} reconsider spherically symmetric space-times, but now in general spherical polar {coordinates} rather than quasi-Cartesian coordinates. {In summary, this} article provides a set of general-{purpose calculational} tools {that can readily be adapted} for mimicking various interesting {(curved)} space-times by using nontrivial susceptibility tensors in general {(background-flat)} laboratory settings.
}


\def\dif{{\mathrm{d}}}
\def\O{{\mathcal{O}}}

\def\tr{{\mathrm{tr}}}
\newcommand{\oo}[1]{\frac{1}{#1}} 
\renewcommand*{\arraystretch}{1.2} 
\definecolor{purple}{rgb}{1,0,1}
\definecolor{lime}{HTML}{A6CE39} 

\begin{document}
\section{Introduction}\label{S:intro}

In three
 earlier articles~\cite{Schuster:2017, Schuster:2018, Schuster:2018b}, {the present authors} have carefully reanalyzed and re-explored the notion of electromagnetic analogue space-times. {In all of these metamaterial analogues, we were very careful to distinguish the background (flat) laboratory metric from the effective (curved space-time) analogue metric.}

In the first article~\cite{Schuster:2017}, we readdressed the {one-century-old} question of just when a (possibly moving) electromagnetic medium is equivalent (in a purely algebraic sense \cite{Schuster:2018b}) to an \emph{effective} 
 Lorentzian metric \cite{Gordon}.
In a simplified setting (static with appropriate coordinate choices), Landau and Lifschitz~\cite{Landau-Lifshitz}, in their influential book ``The classical theory of fields'', write (\S 90, p. 257 of the 3rd edition, 1971):
\begin{quote}
The reader should note the analogy (purely formal, of course) of [the Maxwell equations in a gravitational field]  to the Maxwell equations for the electromagnetic field in material media. In particular, in a static gravitational field [the constitutive relations] {reduce to} 
 $\mathbf{D} = \mathbf{E}/\sqrt{h}$, $\mathbf{B} = \mathbf{H}/\sqrt{h}$. We may say that with respect to its effect on the electromagnetic field a static gravitational field plays the role of a medium with electric and magnetic permeabilities $\epsilon=\mu = 1/\sqrt{h}$.
\end{quote}
In the discussion below, we shall go far beyond this simplified setting, allowing for generic gravitational fields.
That is, the electromagnetic medium is taken to be characterized by effectively three-dimensional permittivity $\epsilon$, permeability $\mu$, magnetoelectric $\zeta$ tensors, and 4-velocity $V^a$.
We used these four quantities to construct an \emph{{effective}} Lorentzian metric---an analogue space-time. 
Specifically, we explicitly constructed (up to an undetermined conformal factor) the effective space-time metric in terms of the optical tensors and \emph{{vice versa}}. 

In the second article~\cite{Schuster:2018}, we explicitly constructed the flat-space laboratory (Cartesian coordinate) susceptibility tensors appropriate for mimicking the Schwarzschild geometry in various quasi-Cartesian forms. (Specifically, we analyzed curvature, isotropic, Kerr--Schild, Painlev\'e--Gullstrand, and Gordon forms of the Schwarzschild geometry.)
We also analyzed general static spherically symmetric space-times, and the Kerr geometry~\cite{Kerr1,Kerr2,Kerr-book,Kerr-intro}, in both Kerr--Schild and Doran forms~\cite{Doran,River} (both of which are easily put into quasi-Cartesian form~\cite{Schuster:2018}). 

In the third article \cite{Schuster:2018b}, we contrasted the essentially algebraic nature of the present electromagnetic analogue space-time model with other potential electromagnetic analogues, which would be more in line with the traditional, analytic approach: Instead of building on tensorial similarities in the action (see the next section), one would traditionally look for a specific differential equation (usually the Laplace--Beltrami equation on a curved background) for making comparisons between electromagnetism and gravitation. This also allowed us to demonstrate simple cartographic effects on the analogue Hawking temperature that are to be expected from simple \emph{{spatial}} rescaling. 
This furthermore links back to existing work on cartography in electromagnetic space-time {analogues} 
 \cite{Thompson:2015,Fathi:2016}.

Unfortunately, working with the Kerr geometry (or more generally, arbitrary axially symmetric space-times) in Boyer--Lindquist coordinates is technically much messier {than working in Kerr--Schild and Doran forms, or in spherical symmetry}. {The difficulty arises} because there is no longer any natural unit-determinant quasi-Cartesian form for the metric, and we shall turn to this topic in the current article.

One particular reason that it is desirable to work with 
space-times in Boyer--Lindquist form is that there is then only one off-diagonal metric component. In contrast, dealing with Kerr--Schild or Doran forms, there is {an explicit} trade-off: While Kerr--Schild or Doran forms can easily be put in quasi-Cartesian form (so that the background {(flat)} laboratory metric can be put in Cartesian form), they have multiple off-diagonal elements in the metric. 

Therefore, the Boyer--Lindquist form minimizes the number of off-diagonal components {in the effective analogue metric}, at the cost of making the background laboratory metric trickier to deal with. {The background laboratory metric is still Riemann flat, just presented in unusual coordinates.}  (Some parts of the calculation {to be presented} below are much simpler than the general discussion in~\cite{Schuster:2017}, or the quasi-Cartesian discussion in~\cite{Schuster:2018}; other parts of the discussion are considerably more subtle.)

After {first} dealing with {generic} Boyer--Lindquist space-times, we turn to arbitrary space-times, not even necessarily stationary, and perform a similar analysis. While the background laboratory metric is always taken to be Riemann flat, we {will now focus on} arbitrary curvilinear coordinate systems. 

Finally, we {shall} reanalyze static spherically symmetric space-times eschewing quasi-Cartesian coordinates. Whereas quasi-Cartesian coordinates are particularly useful for discussing laboratory physics, certain {purely} theoretical computations can be more clearly carried out using spherical polar coordinates{. This} means that one has to keep track of some scalar densities arising from metric determinants (both effective {metric} and background {metric determinants}) and be much more careful when raising and lowering indices. 

\textls[-15]{For further background on these topics{,} see our three earlier articles~\cite{Schuster:2017,Schuster:2018, Schuster:2018b}, the very early 1923 article by  Gordon~\cite{Gordon}, {and the early discussion by Landau and Lifshitz~\cite{Landau-Lifshitz}}. Other relevant{, relatively early} articles come from the general relativity community~\cite{Pham,Plebanski:1960,Plebanski:1970,deFelice:1971,Skrotskii,Balzas, Anderson,HehlObukhov:2003}.  {More recently, the optics community has developed the closely related notion of ``transformation optics'', leading to an extensive body of work~\cite{Joerg:2010,Joerg:2011a,Joerg:2011b,Thompson:2015,Fathi:2016,Thompson:2017,Sch-int,Leonhardt:1999}.}
The electromagnetic analogue space-times are a natural complement to the acoustic analogue space-times  of~\cite{Unruh:1980,Visser:1993,Visser:1997,Barcelo:2000,Barcelo:2001,Barcelo:2003,Visser:2010,Fischer:vortex,Silke:massive-KG,quasiparticle}. For a much more general background and history, see~\cite{LRR,Visser:2001,Visser:2013,Cropp:2015,Liberati,Vorticity}.}

\section{\bf General Strategy}

	Consider electromagnetism in curved space-time. Let the coordinates be $x^a$ and take the effective metric to be $g_{ab}$. We shall explicitly write the contravariant inverse effective metric as $[g^{-1}]^{ab}$. 
Then the curved space-time electromagnetic Lagrangian is 
\cite{Schuster:2017,Schuster:2018b}:
\begin{equation}
	S_{\text{space-time}} =  \oo{8} \int \sqrt{-\det(g_{ab})} 
	\left([g^{-1}]^{ac}  [g^{-1}]^{bd} - [g^{-1}]^{ad} [g^{-1}]^{bc} \right)
	F_{ab} \; F_{cd} \; \dif ^4 x.
\end{equation}
We  wish to mimic this in some laboratory setting, using the same coordinates but now a laboratory (background) metric $(g_0)_{ab}$---which will typically be flat space-time, {but} possibly in unusual coordinates. 
For convenience, we shall assume that the laboratory (background) metric takes the block-diagonal {time+space} form:
\begin{equation}
[g_0]_{ab} = \left[\begin{array}{c|c} 
-1 & 0\\
\hline
0 & [g_0]_{ij}(\xi)\\
\end{array}\right];  \qquad \det([g_0]_{ab}) = - \det([g_0]_{ij}).
\end{equation}
{Then} for a general (linear) metamaterial, 
\begin{equation}
	S_{\text{laboratory}} =\oo{2} \int \sqrt{-\det([g_0]_{ab})} 
	\left( \epsilon^{ij} E_i E_j + 2\zeta^{ij} B_i E_j - [\mu^{-1}]^{ij} B_i B_j \right)
	\dif ^4 x.
\end{equation}
{Equivalently}, to be very explicit,
\begin{equation}
	S_{\text{laboratory}} =\oo{2} \int \sqrt{+\det([g_0]_{ij})} 
	\left( \epsilon^{ij} E_i E_j + 2\zeta^{ij} B_i E_j - [\mu^{-1}]^{ij} B_i B_j \right)
	\dif t \,\dif ^3 \zeta.
\end{equation}
{We} wish to match up these two actions, {$S_{\text{space-time}}$ and $S_{\text{laboratory}}$}, to develop an equivalent laboratory analogue to the curved space-time---at least insofar as it comes to the behavior of electromagnetic fields.
Specifically, we wish to mimic the effective metric $g_{ab}$ (up to some undetermined  conformal factor) by using specified susceptibility tensors $\{ \epsilon^{ij}, [\mu^{-1}]^{ij}, \zeta^{ij} \}$ and a suitable background metric $[g_0]_{ab}$ for laboratory physics. (Conformal invariance of the susceptibility tensors $\{ \epsilon^{ij}, [\mu^{-1}]^{ij}, \zeta^{ij} \}$ is an automatic consequence of the conformal invariance of electromagnetism in (3 + 1) dimensions.
In contrast, the conformal invariance of the Hawking temperature is a somewhat deeper result, applicable not just to electromagnetism~\cite{Jacobson}.)
Indices will always be raised and lowered using the laboratory metric $g_0$ (which is why we need to use the {explicit} notation $[g^{-1}]^{ab}$  for the inverse of the effective metric we want to mimic).

We now consider the constitutive tensor~\cite{Schuster:2017}:
\begin{equation}
Z^{abcd} = \oo{2}\;\sqrt{\frac{\det(g_{ab})}{\det([g_0]_{ab})}} \;\; 
\left([g^{-1}]^{ac}  [g^{-1}]^{bd} - [g^{-1}]^{ad} [g^{-1}]^{bc} \right),
\label{eq:Z-general}
\end{equation}
which mimics the electromagnetic properties of the effective metric $g_{ab}$. 
(Note that $Z^{abcd} $ is, {by construction}, a true tensor, not a tensor density.)
When no confusion can arise, we simplify $\det(g_{ab})\to\det(g)$ and $\det([g_0]_{ab})\to \det(g_0)$. 
The laboratory permittivity, permeability, and magnetoelectric tensors $\{ \epsilon^{ij}, [\mu^{-1}]^{ij}, \zeta^{ij} \}$ are then~\cite{Schuster:2017}:
\begin{equation}
\epsilon^{ij} =  -2 \, Z^{i0j0}; 
\qquad 
[\mu^{-1}]^{ij} =  \oo{2}\;\varepsilon^i{}_{kl}\,\varepsilon^j{}_{mn}\,Z^{klmn};
\qquad 
\zeta^{ij} = \varepsilon^i{}_{kl}\,Z^{klj0}.
\end{equation}
{The} $\varepsilon^i{}_{kl}$ appearing here have to be interpreted as three-dimensional Levi--Civita pseudo-tensors defined in terms of the background spatial 3-metric $[g_0]_{ij}$ and its metric density $\sqrt{\det([g_0]_{ij})}$, with indices raised and lowered using the background spatial 3-metric. This guarantees that $\epsilon^{ij}$ and $[\mu^{-1}]^{ij}$ are true $T^2_0$ tensors under spatial coordinate transformations, while $\zeta^{ij}$ is a pseudo-tensor under parity reversal. (The computations are much simpler if the laboratory metric is Cartesian and the space-time metric quasi-Cartesian; see reference~\cite{Schuster:2018}).
The net result of this procedure is that
\begin{equation}
Z^{abcd} = Z^{abcd}(\varepsilon^{ij}, [\mu^{-1}]^{ij}, \zeta^{ij}, [g_0]_{ij})
\end{equation}
can be written as an explicit function of the optical tensors, while conversely, the optical tensors
\begin{equation}
\{ \epsilon^{ij}, [\mu^{-1}]^{ij}, \zeta^{ij} \} = f( g_{ab}, [g_0]_{ab})
\end{equation}
can be written as explicit functions of the effective metric $g_{ab}$ and background metric $[g_0]_{ab}$.
This implies a stringent compatibility condition (linking the permittivity, permeability, and magnetoelectric tensors) that must be satisfied in order for the analogy to work in the sense of Gordon \cite{Gordon}. {For scalar permittivity and permeability, this gives the compatibility condition $\epsilon=\mu$ of the above quote from Landau and Lifshitz~\cite{Landau-Lifshitz}. The tensorial generalization to $\epsilon^{ij}=\mu^{ij}$ was then rederived several times in the literature with different inspirations \cite{deFelice:1971,Rubilar:2002vfs,Balakin:2005dk}. Even the most general cases can always (at least locally) be 
reduced to this case by appropriate frame changes \cite{Schuster:2017}.} This is the general algorithm---some of the intermediate stages in the specific computations below are somewhat tedious---but the final results are clean, useful, and easy to interpret.

\section{Boyer--Lindquist Space-Times} 

{Essentially, all} physically interesting stationary axisymmetric space-times (not just  Kerr or Kerr--Newman space-times) can be put into Boyer--Lindquist form. Adopting quasi-spherical-polar coordinates, with the coordinates ordered as $(t,r,\theta,\phi)$, with $\theta$ and $\phi$ having the usual interpretation as polar and azimuthal angles, respectively,  it is sufficient to note that without significant loss of generality, any  stationary axisymmetric geometry can be put in the 
form~\cite{Wald,MTW}:
\begin{equation}
g_{ab} = \left[\begin{array}{c|cc|c} 
g_{tt}  & 0 & 0& g_{t\phi}\\
\hline
0 & g_{rr} &0 &0\\
0&0&g_{\theta\theta}&0\\
\hline
g_{t\phi}&0&0&g_{\phi\phi}
\end{array}\right].
\end{equation}
{Here,} the metric components depend only on $(r,\theta)$ and are independent of $(t,\phi)$. 
The inverse metric is easily computed{:}
\begin{equation}
[g^{-1}]^{ab} = 
\left[\begin{array}{c|cc|c} 
g_{\phi\phi}/ g_2  & 0 & 0& - g_{t\phi} /g_2 \\
\hline
0 & 1/g_{rr} &0 &0\\
0&0&1/g_{\theta\theta}&0\\
\hline
-g_{t\phi} /g_2 &0&0&g_{tt} /g_2
\end{array}\right]
=\begingroup
\renewcommand*{\arraystretch}{1.5}
\left[\begin{array}{c|c} 
[g^{-1}]^{00} & {[g^{-1}]^{0j}}\\
\hline
{[g^{-1}]^{0i}} & [g^{-1}]^{ij}
\end{array}\right]. \endgroup
\end{equation}
l{Here,} $g_2 = g_{tt} \,g_{\phi\phi}-g_{t\phi}^2$, and $\det(g_{ab})= g_2\,g_{rr}\,g_{\theta\theta}$.
Note that $g_{tt}=0$ at any ergo-surface, while it is easy to convince oneself that $g_2=0$ at any horizon; equivalently, $g_{rr}=\infty$ at any horizon. (See, for instance, references~\cite{Kerr-book,Kerr-intro}.)

The ``laboratory'' will be assumed to be flat Minkowski space, in coordinates that we will label $(t,r,\theta,\phi) = (t,\xi^i)$.
We assume that the background metric takes the following block-diagonal form:
\begin{equation}
[g_0]_{ab} = \left[\begingroup
\renewcommand*{\arraystretch}{1.5}\begin{array}{c|c} 
-1 & 0 \\\hline
0 & {[g_0]_{ij}(\xi)}\\
\end{array}\endgroup\right],  \qquad \det([g_0]_{ab}) = - \det([g_0]_{ij}).
\end{equation}
{Here,} $[g_0]_{ij}(\xi)$ is the 3-metric corresponding to some (essentially arbitrary) coordinate representation of flat 3-space. For instance, these coordinates might be spherical polar, oblate spheroidal, prolate spheroidal, or cylindrical coordinates (in which case, one would more likely relabel $\theta\to z$), or something even more exotic (for example, parabolic cylindrical, paraboloidal, elliptic cylindrical, ellipsoidal, bipolar, toroidal, conical, or general orthogonal coordinates). 
Let us now explicitly calculate the laboratory permittivity, permeability, and magnetoelectric tensors $\{ \epsilon^{ij}, [\mu^{-1}]^{ij}, \zeta^{ij} \}$.

\subsection{Permittivity Tensor} 
We start by noting the following:
\begin{align}
\epsilon^{ij} &=  -\sqrt{ \frac{\det(g)}{\det(g_0)} } \; 
\left( [g^{-1}]^{ij}  [g^{-1}]^{00}- [g^{-1}]^{0i}  [g^{-1}]^{0j}  \right)
\nonumber\\
&= -\sqrt{\frac{\det(g)}{\det(g_0)}} \; \oo{g_2} \; \left[\begin{array}{cc|c} 
 g_{\phi\phi}/ g_{rr} &0 &0\\
0&g_{\phi\phi}/ g_{\theta\theta}&0\\
\hline
0&0&1
\end{array}\right]^{ij}.
\end{align}
{Now} in the physically interesting region (in the domain of outer communication, outside the outermost noncosmological horizon), both $\det(g)$ and $g_2$ are negative, while $\det(g_0)$ is negative everywhere. This allows us to rewrite $\epsilon^{ij}$ as:
\begin{equation}
	\epsilon^{ij} = \sqrt{\frac{g_{rr}g_{\theta\theta}}{g_2 \det(g_0)}} 
	\left[\begin{array}{cc|c} 
	 g_{\phi\phi}/ g_{rr} &0 &0\\
	0&g_{\phi\phi}/ g_{\theta\theta}&0\\
	\hline
	0&0&1
	\end{array}\right]^{ij}.
\end{equation}
{Note} that this is manifestly conformally invariant (under conformal rescaling of the effective {space-time} metric $g_{ab}$ one is mimicking) as it must be (due to the conformal invariance of electromagnetism in (3 + 1) dimensions); see, for instance, references~\cite{Schuster:2017,Schuster:2018}. 

\subsection{Permeability Tensor} 

We start by noting the following:
\begin{align}
	[\mu^{-1}]^{ij} &=  \oo{2}\;\varepsilon^i{}_{kl}\,\varepsilon^j{}_{mn}\,Z^{klmn}
	\nonumber\\
	&=\oo{2}\;\varepsilon^i{}_{kl}\,\varepsilon^j{}_{mn} \sqrt{\frac{\det(g)}{\det(g_0)}} \; 
	\left([g^{-1}]^{km}  [g^{-1}]^{ln} \right).
\end{align}
{Now} in terms of the Levi--Civita tensor density $\bar\varepsilon_{ijk} = \hbox{signum}(ijk)$, we {can write the true tensor identities:}
\begin{align}
\varepsilon_{ijk}  &= \sqrt{-\det(g_0)} \; \bar\varepsilon_{ijk},
\end{align}
and
\begin{align}
\varepsilon^i{}_{jk}  &= [g_0]^{ip} \sqrt{-\det(g_0)} \; \bar\varepsilon_{pjk},
\end{align}
whence
\begin{equation}
[\mu^{-1}]^{ij} =\oo{2}\; [g_0]^{ip} [g_0]^{jq}  [-\det(g_0)] \sqrt{\frac{\det(g)}{\det(g_0)}} \; \bar\varepsilon_p{}_{kl}\,\bar\varepsilon_q{}_{mn} 
\left([g^{-1}]^{km}  [g^{-1}]^{ln} \right).
\end{equation}
{Now} we know that $ [g^{-1}]^{ln}$ is diagonal, so $\bar\varepsilon_p{}_{kl}\,\bar\varepsilon_q{}_{mn} 
\left([g^{-1}]^{km}  [g^{-1}]^{ln} \right)$ is also {guaranteed to be} diagonal. Then,
\begin{equation}
[\mu^{-1}]^{ij} = [g_0]^{ip} [g_0]^{jq}  \sqrt{\det(g) \det(g_0)} 
\left[\begin{array}{cc|c} 
g_{tt}/(g_{\theta\theta} g_2)&  0 &0\\
0&g_{tt}/(g_{rr} g_2) & 0\\
\hline
0&0&1/(g_{rr} g_{\theta\theta})
\end{array}\right]_{pq}\!\!\!.
\end{equation}
{(Note }in passing the conformal invariance under rescaling of the effective metric $g_{ab}$.)

Matrix inversion (and raising indices using the background metric $g_0$) now yields:
\begin{equation}
\mu^{ij} =  \sqrt{\frac{g_{rr} g_{\theta\theta}}{g_2 \det(g_0)}}
\left[\begin{array}{cc|c} 
g_2/(g_{tt} g_{rr})&  0 &0\\
0&g_2/(g_{tt} g_{\theta\theta}) & 0\\
\hline
0&0&1
\end{array}\right]^{ij}.
\end{equation}
{Note} that this is conformally invariant (under conformal rescaling of the effective metric $g_{ab}$ one is mimicking) as it must be (due to the conformal invariance of electromagnetism in (3 + 1) dimensions); see, for instance, references~\cite{Schuster:2017,Schuster:2018}.

\subsection{Magnetoelectric Tensor} 

We start by noting
\begin{equation}
\zeta_{\,i}{}^j = -\oo{2}\; \sqrt{\frac{\det(g)}{\det(g_0)}} \;  \left(  \varepsilon_{i}{}_{kl}[g^{-1}]^{0l} [g^{-1}]^{kj}  \right).
\end{equation}
{Inserting} the explicit expressions for $[g^{-1}]^{0l}$ and $\varepsilon_{i}{}_{k\phi}$ and simplifying gives 
\begin{equation}
\zeta_i{}^j = \oo{2}\; g_{t\phi} \sqrt{\frac{- g_{rr} g_{\theta\theta}}{g_2}} 
\left[\begin{array}{cc|c} 
0 & 1/g_{\theta\theta} &0\\
-1/g_{rr}&0&0\\
\hline
0&0&0
\end{array}\right]_i^{\;\;j}.
\end{equation}
{Note} again the conformal invariance under rescaling of the effective metric $g_{ab}$ that one is trying to mimic.
{Note further that the magnetoelectric tensor is proportional to the only off-diagonal metric component $g_{t\phi}$. Consequently, magnetoelectric effects vanish in the absence of rotation.}

\subsection{Summary for Metrics in Boyer--Lindquist Coordinates} 

Collecting results, for Boyer--Lindquist space-times, we have the following:
\begin{equation}
\epsilon^{ij} = \sqrt{\frac{g_{rr}g_{\theta\theta}}{g_2 \det(g_0)}} 
\left[\begin{array}{cc|c} 
 g_{\phi\phi}/ g_{rr} &0 &0\\
0&g_{\phi\phi}/ g_{\theta\theta}&0\\
\hline
0&0&1
\end{array}\right]^{ij}.
\end{equation}
\begin{equation}
\mu^{ij} =   \sqrt{\frac{g_{rr} g_{\theta\theta}}{g_2 \det(g_0)}}
\left[\begin{array}{cc|c} 
g_2/(g_{tt} g_{rr})&  0 &0\\
0&g_2/(g_{tt} g_{\theta\theta}) & 0\\
\hline
0&0&1
\end{array}\right]^{ij}.
\end{equation}
\begin{equation}
\zeta_i{}^j = \oo{2}\; g_{t\phi} \sqrt{\frac{- g_{rr} g_{\theta\theta}}{g_2}} 
\left[\begin{array}{cc|c} 
0 & 1/g_{\theta\theta} &0\\
-1/g_{rr}&0&0\\
\hline
0&0&0
\end{array}\right]_i^{\;\;j}.
\end{equation}

{While} the {abstract} computation required to get to this stage has been slightly tedious, the final results are fully explicit, and quite general. A number of interesting implications can immediately be read off.

\subsection{Implications for Metrics in Boyer--Lindquist Coordinates}

\begin{itemize}

\item 
First, note that $\epsilon^{\phi\phi}=\mu^{\phi\phi}$.  (This is related to what we saw happened for spherical symmetry in reference~\cite{Schuster:2018}. The general point is that electromagnetic properties in the direction of the 3-vector $[g^{-1}]^{0i}$ are degenerate.) Indeed, all the components of the permittivity tensor $\epsilon^{ij}$ are well defined down to {the} outermost noncosmological horizon ($g_2=0$, $g_{rr}=\infty$).

\item
Second, note that while $\mu^{\phi\phi}$ is well defined all the way to the outermost noncosmological horizon, $\mu^{rr}$ and $\mu^{\theta\theta}$ are only well defined down to the outermost noncosmological  ergo-surface (where $g_{tt}=0$). 

\item
Third, note that the magnetoelectric tensor $\zeta^{ij}$ is well defined down to the outermost noncosmological  horizon ($g_2=0$, $g_{rr}=\infty$).

\item
Fourth, note that the difference $\epsilon^{ij} - \mu^{ij}$ is relatively simple:
\begin{equation}
\epsilon^{ij} - \mu^{ij} = 
\sqrt{\frac{g_{rr} g_{\theta\theta}}{g_2 \det(g_0)}}  \; \frac{ g_{t\phi}^2}{g_{tt}}
\left[\begin{array}{cc|c} 
 1/g_{rr} &0 &0\\
0&1/g_{\theta\theta}&0\\
\hline
0&0&0
\end{array}\right]^{ij}.
\end{equation}
Observe that as $g_{t\phi}\to0$ (that is, as the rotation is switched off), we find $\epsilon^{ij}=\mu^{ij}$ and $\zeta^{ij} = 0$, the standard compatibility condition for static space-times~\cite{Schuster:2017,Schuster:2018}.

\item
Observe that the magnetoelectric tensor always has a zero-eigenvalue  eigenvector, currently in the $\phi$ direction, and so $\det(\zeta_{\,i}{}^j)=0$ for all Boyer--Lindquist mimics. 

\item
Observe that:
\begin{equation}
(\zeta^2)_i{}^k = \zeta_i{}^j \zeta_j{}^k= \oo{4}\; \frac{g_{t\phi}^2}{g_2} 
\left[\begin{array}{cc|c} 
1 & 0&0\\
0 &1&0\\
\hline
0&0&0
\end{array}\right]_i^{\;\;k}.
\end{equation}
This is actually proportional to a projection operator onto the directions perpendicular to the 3-vector $[g^{-1}]^{0i}$.
(We saw similar things happen in the quasi-Cartesian analysis of reference~\cite{Schuster:2018}.)

\item
Observe that:
\begin{equation}
\tr(\zeta^2) = \zeta_i{}^j \zeta_j{}^i= \oo{2}\; \frac{g_{t\phi}^2}{g_2}.
\end{equation}
This is a nice scalar invariant describing the strength of the magnetoelectric effect, well defined down to the outermost noncosmological horizon (where $g_2=0$).
\item
Finally, note that, for small $g_{t\phi}$, that is, $|g_{t\phi}| \ll \sqrt{|g_{tt} g_{\phi\phi}|}$, we can explicitly write the following:
\begin{equation}
\epsilon^{ij} = - \frac{\sqrt{\det(g)}}{\sqrt{ \det(g_0)}} \; \oo{g_{tt}}
\left[\begin{array}{cc|c} 
 1/ g_{rr} &0 &0\\
0&1/ g_{\theta\theta}&0\\
\hline
0&0&1/g_{\phi\phi}
\end{array}\right]^{ij} \; \left(1 + \mathcal{O}(g_{t\phi}^2) \right) = \mu^{ij} .
\end{equation}
\begin{equation}
\zeta_i{}^j = \oo{2}\; g_{t\phi} \sqrt{\frac{- g_{rr} g_{\theta\theta}}{g_{tt} g_{\phi\phi}}} 
\left[\begin{array}{cc|c} 
0 & 1/g_{\theta\theta} &0\\
-1/g_{rr}&0&0\\
\hline
0&0&0
\end{array}\right]_i^{\;\;j} \left(1 + \mathcal{O}(g_{t\phi}^2) \right).
\end{equation}
\end{itemize}

\section{Specific Concrete Examples}\label{sec:examples}
Let us now consider some specific concrete examples.

\subsection{Kerr}\label{subsec:kerr}
The Kerr space-time represents a rotating vacuum black hole in standard general relativity~\cite{Kerr1,Kerr2,Kerr-book,Kerr-intro}. 
We have already investigated the Cartesian Kerr–Schild and Cartesian Doran forms of the Kerr metric in reference~\cite{Schuster:2018}, but now turn our attention to the Boyer--Lindquist form of the Kerr metric.
Defining as usual 
\begin{equation}
    \Delta= r^2-2mr+a^2; \qquad \Sigma = r^2+a^2\,\cos^2\theta;
\end{equation}
the Kerr line element can be written as ~\cite{Kerr-book,Kerr-intro}:
\begin{eqnarray}
    \dif s^2 &=& - \left[1-\frac{2mr}{\Sigma}\right]\dif t^2 
    - \frac{4mar \sin^2\theta}{\Sigma} \dif \phi \dif t 
    +\left[r^2+a^2+\frac{2ma^2r\sin^2\theta}{\Sigma}\right]\sin^2\theta\dif\phi^2
    \nonumber\\
    &&
    + \frac{\Sigma}{\Delta} \dif r^2 
    + \Sigma \dif\theta^2.
\end{eqnarray}
{For} convenience, we can explicitly take the background metric to be the $m=0$ limit of Kerr, which is known to be {Riemann-flat Minkowski space in oblate spheroidal coordinates}~\cite{Kerr-book,Kerr-intro}:
\begin{eqnarray}
    (\dif s^2)_0 &=& - \dif t^2 
    +\left[r^2+a^2\right]\sin^2\theta\dif\phi^2
    + \frac{\Sigma}{r^2+a^2} \dif r^2 
    + \Sigma \dif\theta^2.
\end{eqnarray}
{Here,} the free parameter $a$ specifies the distance of the oblate spheroid's two foci from the origin, and for $a\to0$, one regains spherical coordinates with a single focus at the origin.
Then,
\begin{equation}
    g_2\to - \Delta \; \sin^2\theta; 
    \qquad 
    \det(g_{ab}) \to -\Sigma^2 \; \sin^2\theta;
    \qquad 
    \det([g_0]_{ab}) \to -\Sigma^2 \; \sin^2\theta.
\end{equation}
Furthermore,
\begin{equation}
     \sqrt{\frac{g_{rr} g_{\theta\theta}}{g_2 \det(g_0)}} \to 
     \oo{\Delta \sin^2\theta}; \qquad
     \sqrt{\frac{-g_{rr} g_{\theta\theta}}{g_2}} \to 
     \frac{\Sigma}{\Delta \sin\theta}.
\end{equation}
Collecting results, for Kerr space-time, we have the following:
\begin{equation}
\epsilon^{ij} = \oo{\Delta\Sigma}
\left[\begin{array}{cc|c} 
 \Delta\left[r^2+a^2+\frac{2ma^2r\sin^2\theta}{\Sigma}\right]
 &0 &0\\
0&\left[r^2+a^2+\frac{2ma^2r\sin^2\theta}{\Sigma}\right]&0\\
\hline
0&0&\frac{\Sigma}{\sin^2\theta}
\end{array}\right]^{ij},
\end{equation}
and
\begin{equation}
\mu^{ij} =  
\left[\begin{array}{cc|c} 
\frac{\Delta}{\Sigma-2mr}&  0 &0\\
0&\oo{\Sigma-2mr} & 0\\
\hline
0&0&\oo{\Delta\sin^2\theta}
\end{array}\right]^{ij}.
\end{equation}

Finally, for the magnetoelectric tensor, we have
\begin{equation}
\zeta_i{}^j = \frac{mar\sin\theta}{\Sigma} 
\left[\begin{array}{cc|c} 
0 & 1/\Delta &0\\
-1&0&0\\
\hline
0&0&0
\end{array}\right]_i^{\;\;j}.
\end{equation}

\subsection{Lense--Thirring}\label{subsec:Lense-Thirring}

\def\O{{\mathcal{O}}}
The Lense--Thirring slow rotation approximation to Kerr has a long and complex history in its own right~\cite{Lense-Thirring,Lense-Thirring-translation,Pfister}. However, for current purposes, it is sufficient to take $ a < m \ll r$ and Taylor expand. 
Explicitly keeping the zeroth-order and first-order terms in $a$, one has the following:
\begin{equation}
    \dif s^2 = -(1-2m/r)\dif t^2 +\frac{\dif r^2}{1-2m/r} +r^2(\dif\theta^2 + \sin^2\theta\dif\phi^2) - \frac{4ma\sin^2\theta}{r} \dif \phi \dif t + \O(a^2),
\end{equation}
with the electromagnetic tensors simplifying to the following:
\begin{equation}
\epsilon^{ij} = 
\left[\begin{array}{cc|c} 
1 &0 &0\\
0&\oo{r^2(1-2m/r)} &0\\
\hline
0&0&\oo{r^2(1-2m/r)\sin^2\theta}
\end{array}\right]^{ij} +\O(a^2)= 
\frac{g^{ij}}{1-2m/r} +\O(a^2).
\end{equation}
\begin{equation}
\mu^{ij} = 
\left[\begin{array}{cc|c} 
1 &0 &0\\
0&\oo{r^2(1-2m/r)} &0\\
\hline
0&0&\oo{r^2(1-2m/r)\sin^2\theta}
\end{array}\right]^{ij} +\O(a^2)= 
\frac{g^{ij}}{1-2m/r} +\O(a^2).
\end{equation}
\begin{equation}
\zeta_i{}^j = \frac{ma\sin\theta}{r} 
\left[\begin{array}{cc|c} 
0 & \oo{r^2(1-2m/r)} &0\\
-1&0&0\\
\hline
0&0&0
\end{array}\right]_i^{\;\;j} + \O(a^2).
\end{equation}
{That} is,
\begin{equation}
\zeta_{ij} = \frac{ma\sin\theta}{r} 
\left[\begin{array}{cc|c} 
0 & 1 &0\\
-1&0&0\\
\hline
0&0&0
\end{array}\right]_{ij} + \O(a^2).
\end{equation}
{Note} that this is just what one would expect for Schwarzschild (in curvature coordinates), plus a first-order magnetoelectric effect to mimic slow rotation. {Additionally, note that the naming conventions of electromagnetism and its constitutive tensors differ here compared with the conventions commonly adopted in gravitomagnetism; the closeness of the Lense--Thirring metric to the latter warrants an extra warning concerning this point.}

\section{{Going beyond} Boyer--Lindquist: Arbitrary Space-Times}\label{sec:beyond}

A surprise for arbitrary space-times is that the usual ADM formalism proves to be less than useful, whereas the ``threaded'' (Kaluza--Klein--inspired) form of the metric is more amenable to developing a metamaterial analogue. 
To now proceed beyond Boyer--Lindquist space-times, we first write the metric to be mimicked in threaded form~\cite{Boersma:1994,Bejancu:2015,Gharechahi:2018} (also known as  Kaluza--Klein form~\cite{Schuster:2017,Schuster:2018}):
 \begin{equation}
[g^{-1}]^{ab} =  \left[\begin{array}{cc}
-\alpha^{-2} + \gamma^{-1}_{kl} \beta^k \beta^l & \beta^j\\ 
\beta^i & \gamma^{ij} 
\end{array}\right];  \qquad \det(g)=-\alpha^2\det(\gamma^{-1}).
\label{eq:g-KK1}
 \end{equation}
Equivalently, for the covariant metric, one has:
\begin{equation}
[g]_{ab} =  \left[\begin{array}{cc}
-\alpha^{2}  & \alpha^2 \,\beta_j\\ 
\alpha^2 \,\beta_i & [\gamma^{-1}]_{ij} -\alpha^2\,\beta_i\,\beta_j
\end{array}\right];  \qquad \beta_i = [\gamma^{-1}]_{ik} \,\beta^k.
\label{eq:g-KK1b}
\end{equation}
Note that the entries of the covariant and contravariant metrics are exactly swapped compared with the ADM formalism, hence the ``duality'' of Kaluza--Klein and ADM.
  
This is not at all a restriction on the metric, merely a convenient way of writing it. (We have not enforced the unimodular condition of reference~\cite{Schuster:2018} since we are now not using quasi-Cartesian coordinates, and the unimodular condition is now neither necessary nor useful.)
Note that because the analysis below is purely algebraic, 
there is no need to demand stationarity (time independence)---such an assumption may be convenient, but it is not necessary.

\subsection{Permittivity Tensor} 
We start by noting the following:
\begin{align}
	\epsilon^{ij} &=  -\sqrt{\frac{\det(g)}{\det(g_0)}} \; 
	\left( [g^{-1}]^{ij}  [g^{-1}]^{00}- [g^{-1}]^{0i}  [g^{-1}]^{0j}  \right)
	\\
	&= + \sqrt{\frac{\det(g)}{\det(g_0)}}  
	 \left\{(\alpha^{-2} - \gamma^{-1}_{kl} \beta^k \beta^l )\, \gamma^{ij} + \beta^i \beta^j\right\}.
\end{align}
We can write this as:
\begin{equation}
\epsilon^{ij}  = \oo{\alpha \sqrt{|\det(g_0)|\det(\gamma^{pq})} } 
\left\{(1 - \alpha^2 [\gamma^{-1}]_{kl} \beta^k \beta^l )\, \gamma^{ij} + \alpha^2 \beta^i \beta^j\right\}
\end{equation}
Note that this is a true tensor equation (under arbitrary spatial coordinate transformations), and that it is conformally invariant (under conformal rescaling of $g_{ab}$, the effective metric to be mimicked).

\subsection{Permeability Tensor} 
We start by noting the following:
\begin{align}
	[\mu^{-1}]^{ij} &=  \oo{2}\;\varepsilon^i{}_{kl}\,\varepsilon^j{}_{mn}\,Z^{klmn}
	\\
	&=\oo{2}\; [g_0]^{ip} [g_0]^{jq}  [-\det(g_0)] \sqrt{\frac{\det(g)}{\det(g_0)}} \; \bar\varepsilon_p{}_{kl}\,\bar\varepsilon_q{}_{mn} 
\left(\gamma^{km}  \gamma^{ln} \right).
\end{align}
Lowering indices (using the background metric),
\begin{equation}
[\mu^{-1}]_{ij} =\oo{2}\; [-\det(g_0)] \sqrt{\frac{\det(g)}{\det(g_0)}} \; \varepsilon_i{}_{kl}\,\varepsilon_j{}_{mn} 
\left(\gamma^{km}  \gamma^{ln} \right).
\end{equation}
Now purely as a matter of algebra, for $3\times3$ matrices, we have 
\begin{equation}
\bar\varepsilon_i{}_{kl}\,\bar\varepsilon_j{}_{mn}  \left(\gamma^{km}  \gamma^{ln}\right) =
2 \det(\gamma^{pq}) [\gamma^{-1}]_{ij}.
\end{equation}
{Using} this, together with a matrix inversion, now yields 
\begin{equation}
\mu^{ij} =\; \oo{\alpha \sqrt{|\det(g_0)|\det(\gamma^{pq})} } \;  \gamma^{ij}.
\end{equation}
{Note} that this is a true tensor equation (under arbitrary spatial coordinate transformations), and that it is conformally invariant (under conformal rescaling of $g_{ab}$, the effective metric to be mimicked).

\subsection{Magnetoelectric Tensor} 

Lowering the first index, the best we can do in full generality is (after a very short calculation) as follows:
\begin{equation}
\zeta_{\,i}{}^j = -\oo{2}\; \sqrt{\frac{\det(g)}{\det(g_0)}} \;  \left(  \varepsilon_{i}{}_{kl}\beta^l \gamma^{kj}  \right).
\end{equation}
Note again the conformal invariance under rescaling of $g_{ab}$.
Furthermore, this is a true pseudo-tensor equation (under spatial coordinate transformations).

\subsection{Summary for Arbitrary Space-Times} 

Collecting results, for a generic space-time, we have the following:
\begin{equation}
\epsilon^{ij}  = \oo{\alpha \sqrt{|\det(g_0)|\det(\gamma^{pq})} } 
\left\{(1 - \alpha^2 \gamma^{-1}_{kl} \beta^k \beta^l )\, \gamma^{ij} + \alpha^2 \beta^i \beta^j\right\}
\end{equation}
\begin{equation}
\mu^{ij} =\; \oo{\alpha \sqrt{|\det(g_0)|\det(\gamma^{pq})} } \;  \gamma^{ij}.
\end{equation}
\begin{equation}
\zeta_{\,i}{}^j = -\oo{2}\; \sqrt{\frac{\det(g)}{\det(g_0)}} \;  \left(  \varepsilon_{i}{}_{kl}\beta^l \gamma^{kj}  \right).
\end{equation}
Again, getting to this stage has been slightly tedious, but the final results are fully explicit and relatively simple. Several interesting implications immediately follow.

\subsection{Implications for Arbitrary Space-Times}

\begin{itemize}
\item 
First, note that the difference $\epsilon^{ij}  -\mu^{ij}$ is quite simple:
\begin{equation}
\epsilon^{ij}  -\mu^{ij} = - \frac{\alpha}{\sqrt{|\det(g_0)|\det(\gamma^{pq})} } 
\left\{( \gamma^{-1}_{kl} \beta^k \beta^l )\, \gamma^{ij} -  \beta^i \beta^j\right\}.
\end{equation}
Observe that as $\beta^i\to0$, we again find $\epsilon^{ij}=\mu^{ij}$ (and $\zeta^{ij} = 0$), the standard compatibility condition for static space-times~\cite{Schuster:2017,Schuster:2018}.
\item
Second, note that if we define $||\beta||^2 = ( \gamma^{-1}_{kl} \beta^k \beta^l )$ and $\hat \beta^i = \beta^i/||\beta||$, then
\begin{equation}
\epsilon^{ij}  -\mu^{ij} = - \frac{\alpha ||\beta||^2}{\sqrt{|\det(g_0)|\det(\gamma^{pq})} } 
\left\{\gamma^{ij} -  \hat \beta^i \hat\beta^j\right\}.
\end{equation}
Then in the direction of the 3-vector $ [\gamma^{-1}]_{jk} \beta^k$, we have
\begin{equation}
(\epsilon^{ij} -\mu^{ij}) [\gamma^{-1}]_{jk} \beta^k = 0.
\end{equation}
(This is related to what we saw happen for Boyer--Lindquist above (where we found $\epsilon^{\phi\phi}=\mu^{\phi\phi}$), and also for spherical symmetry in~\cite{Schuster:2018}. The general point is that electromagnetic properties in the direction of the 3-vector $\beta^i = [g^{-1}]^{0i}$ are degenerate.)

\item
For the magnetoelectric tensor, observe that $\beta^i \, \zeta_{\,i}{}^j=0$, so the direction $\beta^i$ is again special. This implies that the magnetoelectric tensor always has {a left eigen\-vector with eigenvalue zero}, and $\det(\zeta_{\,i}{}^j)=0$. 
(This is a general property of electromagnetic media with a single light cone. See \cite{HehlObukhov:2003}, p. 282).
The corresponding {right eigen\-vector with eigenvalue zero} is $[\gamma^{-1}]_{jk} \beta^k$. 
In fact, the magnetoelectric tensor can always be written in the form $\zeta_i{}^k = A_{ij} S^{jk}$ with $A_{ij}$ antisymmetric and $S^{jk}$ symmetric \mbox{$3\times3$ matrices,} which makes the existence of these eigenvectors obvious.
\item
Observe that
\begin{equation}
(\zeta^2)_i{}^j = \zeta_i{}^k\; \zeta_k{}^j= \oo{4}\; {\frac{\det(g)}{\det(g_0)}}   
\left(  \varepsilon_{i}{}_{pl}\beta^l \gamma^{pk}  \right)  
\left(  \varepsilon_{k}{}_{mn}\beta^n \gamma^{mj}  \right).
\end{equation}
Hence,
\begin{equation}
\tr(\zeta^2) =  \oo{2}\;\alpha^2 \;
[\gamma^{-1}]_{ij} \beta^i \beta^j.
\end{equation}
This is a simple scalar invariant (under spatial coordinate transformations) describing the strength of the magnetoelectric effect.
\item
If we choose to rotate our coordinate system so that $\beta^i=(0,0,\beta)$, then things simplify somewhat. The eigenvalues of $\zeta_i{}^j$ are then proportional to $(0, +\beta, - \beta)$, and consequently, the eigenvalues of $(\zeta^2)_i{}^j$ are then proportional to $(0, +\beta^2, +\beta^2)$. Indeed, 
\begin{equation}
(\zeta^2)_i{}^j = \frac{\tr(\zeta^2)}{2} 
\left[\begin{array}{cc|c} 
1 & 0& \,*\\
0 &1&  \,*\\
\hline
0&0&0
\end{array}\right]_i^{\;j},
\end{equation}
which is almost a projection operator. 
\end{itemize}

\section{{Revisiting Spherically} Symmetric Space-Times in Spherical Polar Coordinates}\label{sec:polar}
Finally, it is worth revisiting the analysis of reference~\cite{Schuster:2018} for static spherically symmetric space-times,
but now eschewing the use of quasi-Cartesian coordinates, and allowing for time dependence.  While the quasi-Cartesian coordinates of  reference~\cite{Schuster:2018} correspond to Cartesian coordinates for the background metric describing the laboratory, and so are particularly useful for presentational purposes when phrasing laboratory-based questions, sometimes explicit computations are more cleanly carried out in spherical polar coordinates. 

There is, however, a price to be paid: one has to keep track of some scalar densities arising from metric determinants (both effective and background) and be much more careful raising and lowering indices. 

First, let us adopt $(t,r,\theta,\phi)$ coordinates, and write the flat background metric in the following form:
\begin{equation}
(\dif s_0)^2 = - dt^2 +  [R'(r) dr]^2 + R(r)^2\{d\theta^2+\sin^2\theta \, d\phi^2\}.
\end{equation}
Note $\det([g_0]_{ab}) = -R'(r)^2\,R(r)^4\,\sin^2\theta \neq -1$; the background metric is not unimodular. 
This is the most general form of a flat space-time metric com\-pa\-ti\-ble with explicit spherical symmetry in the sense of being based on spherical polar coordinates.
Then for the metric to be mimicked, $g_{ab}$ we can {write without any loss of generality:}
\begin{equation}
(\dif s)^2 = g_{tt} \,dt^2 + 2 g_{tr} \,dt dt + g_{rr} \,dr^2 + R(r)^2\{d\theta^2+\sin^2\theta \, d\phi^2\}.
\end{equation}
Note that we are using the same coordinates in both the laboratory and the metric to be mimicked,
and we are making extensive use of the assumed spherical symmetry. We are also keeping the metric to be mimicked in as general a form as possible---so that we can simultaneously deal with curvature coordinates (where $R(r)=r$), isotropic coordinates (where $g_{rr}=R(r)^2$), or various off-diagonal coordinates (such as Kerr--Schild coordinates, Painlevé--Gullstrand coordinates, or Gordon coordinates~\cite{Liberati,Vorticity,Rosquist}).

To proceed, we again rewrite the metric to be mimicked in threaded form:
 \begin{equation}
[g^{-1}]^{ab} =  \left[\begin{array}{c|c|cc}
-\alpha^{-2} + \beta^2/\gamma & \beta & 0 & 0\\
\hline
\beta & \gamma &0&0\\
\hline
0&0& R^{-2} & 0 \\
0&0&0&R^{-2} (\sin^2\theta)^{-1}
\end{array}\right]. 
\label{eq:g-KK2}
 \end{equation}
As before, this is a mere rewriting of the metric.
Note that 
\begin{equation}
\det(g_{ab}) = - \alpha^2 \gamma^{-1} R(r)^4\,\sin^2\theta \neq -1,
\end{equation}
and 
\begin{equation}
\det(g_{ab}) /\det([g_0]_{ab}) = \alpha^2 \gamma^{-1} R'(r)^{-2}. 
\end{equation}
(We again emphasize that we have not enforced the unimodular condition of~\cite{Schuster:2018} since we are now not using quasi-Cartesian coordinates, and the unimodular condition is now neither necessary nor useful.)

\subsection{Permittivity Tensor} 

We start by noting
\begin{align}
	\epsilon^{ij} &=   + \frac{\sqrt{\gamma}}{\alpha|R'(r)|}
\left[\begin{array}{c|cc}
1 &0 &0\\
\hline
0& (1- \alpha^{2} \beta^2/\gamma)  \gamma^{-1} R^{-2} & 0 \\
0&0&(1- \alpha^{2} \beta^2/\gamma) \gamma^{-1} R^{-2} (\sin^2\theta)^{-1}
\end{array}\right]^{ij}. 
\end{align}
{Note} that this is conformally invariant (under conformal rescaling of $g_{ab}$).
The somewhat ugly $(\sin^2\theta)^{-1}$ factor can be simplified away by going to a local orthonormal basis in the angular coordinates (adopting an orthonormal \emph{{dyad}}, or \emph{{zweibein}}) and writing
\begin{equation}
\epsilon^{\hat i \hat j}  = + \frac{\sqrt{\gamma}}{\alpha|R'(r)|}
\left[\begin{array}{c|cc}
1 &0 &0\\
\hline
0& (1- \alpha^{2} \beta^2/\gamma)  \gamma^{-1} R^{-2}& 0 \\
0&0&(1- \alpha^{2} \beta^2/\gamma) \gamma^{-1} R^{-2}
\end{array}\right]^{\hat i\hat j}. 
\end{equation}

\subsection{Permeability Tensor} 

It is most efficient to recall the result we obtained for general  space-times:
\begin{equation}
\mu^{ij} =\; \oo{\alpha \sqrt{|\det(g_0)|\det(\gamma^{pq})} } \;  \gamma^{ij},
\end{equation}
and to simply unpack the various contributions to obtain the following:
\begin{equation}
\mu^{ij} = + \oo{\alpha\sqrt{\gamma}|R'(r)|}
\left[\begin{array}{c|cc}
\gamma &0 &0\\
\hline
0& R^{-2} & 0 \\
0&0&R^{-2} (\sin^2\theta)^{-1}
\end{array}\right]^{ij}.
\end{equation}
Alternatively, we can rewrite this as 
\begin{equation}
\mu^{ij} = + \frac{\sqrt{\gamma}}{\alpha|R'(r)|}
\left[\begin{array}{c|cc}
1 &0 &0\\
\hline
0& \gamma^{-1} R^{-2} & 0 \\
0&0&\gamma^{-1}  R^{-2}   (\sin^2\theta)^{-1}
\end{array}\right]^{ij}.
\end{equation}
{Note} that this is conformally invariant (under conformal rescaling of $g_{ab}$).
Going to a local orthonormal basis for the angular coordinates, adopting an orthonormal \emph{(dyad/zweibein)}, this simplifies to the following:
\begin{equation}
\mu^{\hat i \hat j} = + \frac{\sqrt{\gamma}}{\alpha|R'(r)|}
\left[\begin{array}{c|cc}
1 &0 &0\\
\hline
0& \gamma^{-1} R^{-2} & 0 \\
0&0&\gamma^{-1} R^{-2} 
\end{array}\right]^{\hat i \hat j} .
\end{equation}

\subsection{Magnetoelectric Tensor} 

We start by noting
\begin{equation}
\zeta_{\,i}{}^j = -\frac{\beta}{2}\; \sqrt{\frac{\det(g)}{\det(g_0)}} \;  \left(  \varepsilon_{i}{}_{kr} \gamma^{kj}  \right)
=
-\frac{\beta}{2}\; \sqrt{\det(g)} \;  \left(  \bar\varepsilon_{i}{}_{kr} \gamma^{kj}  \right).
\end{equation}
That is,
\begin{equation}
\zeta_{\,i}{}^j = -\frac{\beta}{2}\;  \frac{\alpha |R'| }{\sqrt{\gamma}} \;  \;
\left[\begin{array}{c|cc}
0 &0 &0\\
\hline
0 & 0 & (\sin\theta)^{-1} \\
0&-\sin\theta& 0
\end{array}\right]_i^{\;\;j}.
\end{equation}
{Note} again the conformal invariance under rescaling of $g_{ab}$.
Adopting an orthonormal dyad in the angular directions{,} we have the relatively simple form:
\begin{equation}
\zeta_{\,\hat i}{}^{\hat j} = -\frac{\beta}{2}\;  \frac{\alpha |R'| }{\sqrt{\gamma}} \;  \;
\left[\begin{array}{c|cc}
0 &0 &0\\
\hline
0 & 0 & 1 \\
0&-1& 0
\end{array}\right]_{\hat i}^{\;\;\hat j}.
\end{equation}

\subsection{Summary for Spherically Symmetric Space-Times}

Collecting results, for a generic spherically symmetric  space-time we have the following:
\begin{equation}
\epsilon^{\hat i \hat j}  = + \frac{\sqrt{\gamma}}{\alpha|R'(r)|}
\left[\begin{array}{c|cc}
1 &0 &0\\
\hline
0& (1- \alpha^{2} \beta^2/\gamma)  \gamma^{-1} R^{-2}& 0 \\
0&0&(1- \alpha^{2} \beta^2/\gamma) \gamma^{-1} R^{-2}
\end{array}\right]^{\hat i \hat j}. 
\end{equation}
\begin{equation}
\mu^{\hat i \hat j} = + \frac{\sqrt{\gamma}}{\alpha|R'(r)|}
\left[\begin{array}{c|cc}
1 &0 &0\\
\hline
0& \gamma^{-1} R^{-2} & 0 \\
0&0&\gamma^{-1} R^{-2} 
\end{array}\right]^{\hat i \hat j}.
\end{equation}
\begin{equation}
\zeta_{\,\hat i}{}^{\hat j} = -\frac{\beta}{2}\;  \frac{\alpha |R'| }{\sqrt{\gamma}} \;  \;
\left[\begin{array}{c|cc}
0 &0 &0\\
\hline
0 & 0 & 1 \\
0&-1& 0
\end{array}\right]_{\hat i}^{\;\;\hat j}.
\end{equation}
Yet again,
getting to this stage has been slightly tedious, but the final results are fully explicit and relatively easy to work with. Several interesting implications immediately follow.

\subsection{Implications for Spherically Symmetric Space-Times}

\begin{itemize}
\item 
First, note that $\epsilon^{rr} = \mu^{rr}$. This is similar to something that we have seen several times before (the general point being that electromagnetic properties in the direction of the 3-vector $[g^{-1}]^{0i}$ are degenerate). 
\item
Second, note the following: 
\begin{equation}
\epsilon^{\hat i \hat j}  -\mu^{\hat i \hat j} = 
+ \frac{\sqrt{\gamma}}{\alpha|R'(r)|} \frac{\alpha^2\beta^2}{\gamma^2 R^2} 
\left[\begin{array}{c|cc}
0 &0 &0\\
\hline
0& 1& 0 \\
0&0&1
\end{array}\right]^{\hat i \hat j}. 
\end{equation}
Observe that as $\beta^i\to0$, we find $\epsilon^{ij}=\mu^{ij}$ and $\zeta^{ij} = 0$, the standard compatibility condition for static space-times~\cite{Schuster:2017,Schuster:2018}.

\item
Observe that the magnetoelectric tensor always has a zero-eigenvalue eigenvector, now the radial direction, and so $\det(\zeta_{\,i}{}^j)=0$, as required for materials with a single light cone.

\item
Observe that $(\zeta^2)_i{}^j$ is again {proportional to} a projection operator{:}
\begin{equation}
(\zeta^2)_i{}^j = \zeta_i{}^j \zeta_j{}^i= -\oo{4}\;\frac{\alpha^2\beta^2|R'|^2}{\gamma}
\left[\begin{array}{c|cc}
0 &0 &0\\
\hline
0 & 1 & 0 \\
0&0& 1
\end{array}\right].
\end{equation}
Then,
\begin{equation}
\tr(\zeta^2) =  \oo{2}\;\frac{\alpha^2 \beta^2}{\gamma} (R')^2. 
\end{equation}
This is again a simple scalar invariant describing the strength of the magnetoelectric effect.
\end{itemize}

\section{Discussion}\label{sec:d}

When using metamaterial models to mimic interesting general relativistic space-times, there is often a trade{-}off between simplicity of presentation and simplicity of calculation. Certainly, easily accessible experimental laboratories are, for all practical purposes, living in flat space-time, and most typically for presentational purposes, one might like to deal with simple Cartesian coordinates, which underpinned the quasi-Cartesian analysis we carried out in a previous article on bespoke analogue space-times~\cite{Schuster:2018}, wherein the metrics to be mimicked were all cast into unimodular form $\det(g_{ab})=-1$ and raising and lowering indices with the Cartesian background metric was trivial.

In contrast, in the current article, we avoid any quasi-Cartesian assumptions, at the cost of having to deal with and carefully keep track of metric tensor densities (for both the effective metric to be mimicked and the background laboratory metric)---one also has to be careful raising and lowering indices using the background laboratory metric. 
While intermediate stages of the computations were algebraically tedious, the final answers were both tractable and physically interesting.

The reason for going to this extra effort is essentially a theoretical one---some calculations eventually are more tractable in symmetry adapted coordinate systems, though the initial barrier to setting up the formalism is higher. We hope to return to these issues in future work.

\section{Conclusions}\label{sec:c}
{This article provides numerous general-purpose calculational tools for electromagnetically mimicking the metrics (line elements) of curved space-times in terms of $3\times3$ electromagnetic permeability, permittivity, and magnetoelectric tensors. Specifically, given any explicit space-time metric (implicitly making a coordinate choice) and choosing any explicit Riemann-flat laboratory metric (background metric), the permeability, permittivity, and magnetoelectric tensors can be uniquely defined and extracted from the formalism. Several special cases are carefully worked out. }

{Indeed, the} current article deals with three {explicit} themes: Boyer--Lindquist space-times (suitable for dealing with axisymmetric stationary space-times), generic space-times, and finally {(as a consistency check)} spherically symmetric space-times represented in spherical polar coordinates. In all of these situations, we have been able to present quite specific {and easily applicable} formulae specifying the {constitutive} tensors ({the} permittivity, permeability,  and magnetoelectric tensors) required to mimic the given space-time geometry. The long-term goal is to apply these ideas to analogue Hawking radiation~\cite{Reznik:1997}.

\vspace{6pt}

\authorcontributions{Conceptualization, S.S. and M.V.; methodology, S.S. and M.V.; software, S.S. and M.V.; 
validation, S.S. and M.V.; formal analysis, S.S. and M.V.; investigation, S.S. and M.V.; resources, M.V.; 
writing---original draft preparation, S.S. and M.V.;  writing---review and editing,  S.S. and M.V.; 
supervision, M.V.; project administration, M.V.; funding acquisition, M.V. 
All authors have read and agreed to the published version of the manuscript}

\funding{S.S. acknowledges financial support via OP RDE Project No. CZ.02.2.69/0.0/0.0/18\_053/ 0016976 (international mobility of research) and the technical and administrative staff at the Charles University. M.V. was supported by the Marsden Fund via a grant administered by the Royal Society of New Zealand.
}

\dataavailability{All relevant data are reported in the manuscript.} 


\conflictsofinterest{The authors declare no conflicts of interest.}


\begin{adjustwidth}{-\extralength}{0cm}

\reftitle{References and Notes}

\PublishersNote{}
\end{adjustwidth}
\end{document}